\documentclass[12pt,a4paper,final]{iopart}

\usepackage{iopams}
\usepackage{graphicx}
\usepackage{epstopdf}
\usepackage{color}
\usepackage[breaklinks=true,colorlinks=true,linkcolor=blue,urlcolor=blue,citecolor=blue]{hyperref}

\begin{document}
\title[]{Nanostructured optical nanofibres for atom trapping}

\author{M. Daly, V.G. Truong, C.F. Phelan, K. Deasy and S. Nic Chormaic}
\address{Light-Matter Interactions Unit, OIST Graduate University, 1919-1 Tancha, Onna-son, Okinawa 904-0495, Japan}
\ead{sile.nicchormaic@oist.jp}

\begin{abstract}
We propose an optical dipole trap for cold, neutral atoms based on the electric field produced from the evanescent fields in a hollow rectangular slot cut through an optical nanofibre. In particular, we discuss the trap performance in relation to laser-cooled rubidium atoms and show that a far off-resonance, blue-detuned field combined with the attractive surface-atom interaction potential from the dielectric material forms a stable trapping configuration. With the addition of a red-detuned field, we demonstrate how three dimensional confinement of the atoms at a distance of 140 - 200 nm from the fibre surface within the slot can be accomplished.  This scheme facilitates optical coupling between the atoms and the nanofibre that could be exploited for quantum communication schemes using ensembles of laser-cooled atoms.

\end{abstract}

\vspace{2pc}
\section{Introduction}

Evanescent wave devices have been commonly used with atomic systems for the past few decades \cite{ref1,ref2}, but more recently the integration of optical nanofibres, i.e. optical fibres with dimensions smaller than the wavelength of the guided light \cite{ref3,ref4,ref5}, into cold atomic systems \cite{ref6,ref7,ref8} has been the focus of increasing research interest and a comprehensive review of progress is contained in \cite{ref9}. In particular, it has been shown that optical nanofibres can be used for trapping and manipulating cold atoms. Surface traps, such as those presented by Ovchinnikov \emph{et al} \cite{ref10}, adapted for use on curved optical nanofibre surfaces, make use of the evanescent field present when light is guided along thin optical fibres. The induced potential from a red-detuned field produces an attractive potential, while the atom-surface interaction potential is balanced using a blue-detuned potential barrier that prevents atoms from migrating to the fibre surface. In particular, trapping of alkali atoms in the evanescent field surrounding an optical nanofibre using a combination of red- and blue-detuned optical fields, the so-called \emph{two-colour trap}, has been proposed \cite{ref11,ref12,ref13,ref14} and demonstrated \cite{ref15,ref16}, thereby proving the effectiveness of the technique. In these experiments, the trapped atoms are coupled to a mode which is on resonance with the atom transition and exhibit strong coupling and a large optical depth. These features are desirable for quantum information applications and have led to much research on similar nanophotonic systems for trapping and probing atoms \cite{ref17,ref18}. Red-detuned light traps in hollow fibres have also been used to guide atoms \cite{ref19}. Alternative methods for guiding and trapping atoms outside optical nanofibres have been proposed by several groups, but practical implementation is still quite limited. Single colour traps which make use of higher order modes above slab waveguides \cite{ref20}, non-Gaussian beam shapes \cite{ref21}, or mode interference \cite{ref22} to create stable traps have been proposed. Alternative fibre trapping schemes, such as helical trapping potentials around the nanofibre \cite{ref23}, an induced fictitious magnetic field around the fibre \cite{ref24}, or diffracted laser light off the fibre \cite{ref25}, also exist.  All of these recently proposed methods have a common feature in that the atoms are trapped outside the nanofibre, thereby limiting the efficiency of interaction with any light guided by the fibre .

The fabrication of optical nanofibres with very high transmission ($>$99$\%$) for the fundamental mode has become a standard technique \cite{ref26} and alteration of either the chemical or physical properties of optical fibre surfaces is becoming commonplace. For example, the generation of a Bragg grating on an optical nanofibre by selective milling of the surface using a focused ion beam \cite{ref27} and the fabrication of a humidity detector using subwavelength fibres via the application of 80 nm gelatin layers \cite{ref28} have been reported.  More recently, optical nanofibres that permit relatively high transmission of higher order modes have also been fabricated \cite{ref5,ref29} and this opens up the possibility of experimentally investigating a number of heretofore theoretical fibre-based atom trapping schemes.

In this paper, we propose a method to trap and probe atoms inside a rectangular slot cut through the waist region of a silica optical nanofibre. The design of this device is analogous to that of a slot waveguide \cite{ref30}. In a slot waveguide, the slot width is chosen so that it is smaller than the decay length of the evanescent fields. For such structures a high mode confinement between the two waveguides is possible. Mirroring this structure in an optical nanofibre allows us to trap atoms in the slot area, thereby resulting in several advantages over systems where the atoms are trapped outside the fibre. The simplicity of the design also opens up many possibilities for atom trapping, whether through single colour, higher mode trapping, or the addition of more rectangular slots allowing for the creation of spatially localized trapping regions. Here, we focus on a two-colour setup for producing trapping potentials both at the fibre surface and within the slot. By a suitable choice of slot size, deep potentials with substantial trap lifetimes are predicted, with local minima located at positions of 140 nm - 200 nm from the inner surfaces of the slot.

\section{Guided modes of the system}
\subsection{Optical mode distributions}

\begin{figure}[htb!]
 \centering
 \includegraphics[scale=.5]{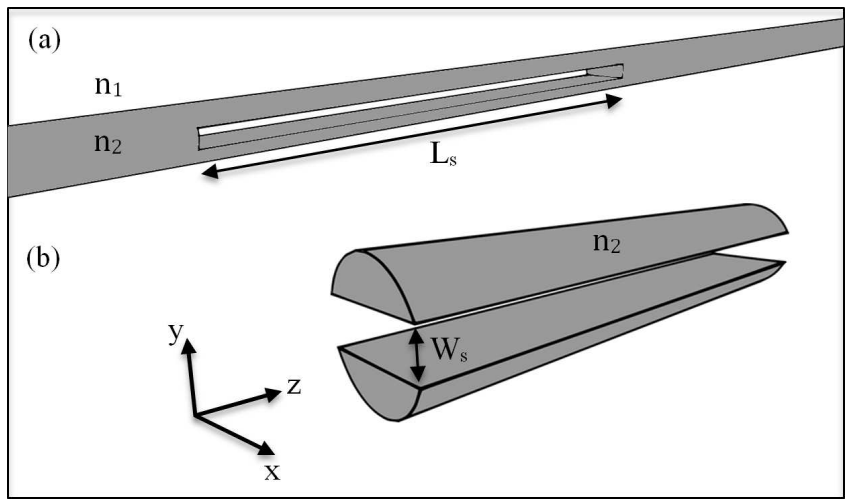}
 \caption{\label{Fig1}Schematic of the slotted fibre for a vacuum-fibre system  $n_1$ = 1 and $n_2$ = 1.45991. (a) Tapered region with slot present, (b) slot region only.}
\end{figure}

The system under study, shown schematically in \Fref{Fig1}, consists of an ultra-thin, vacuum clad, silica nanofibre with a rectangular shaped slot removed from the centre of the fibre. We assume that fibre diameters are of the order of the wavelength of the guided light. We define the axes such that the z-direction corresponds to the fibre axis, the direction parallel to the slot and orthogonal to the z-axis is the \textit{x}-direction, and the remaining axis perpendicular to the slot is the \textit{y}-direction.

For wavelengths of 1064 nm and 720 nm, chosen for their red- and blue-detunings from the 780 nm Rb cooling transition, respectively, and nanofibre diameters ranging from 0.6-1.2  $\mu$m, four distinct guided optical modes can be identified. Using an approach similar to that developed by Anderson \textit{et al} \cite{ref30} the modes are viewed as being either symmetric or anti-symmetric. Furthermore, each mode can propagate with two orthogonal polarisations, giving rise to the four distinct modes.

The supported optical modes of the slotted fibre can be determined from Maxwell'€™s wave equation, equation(\ref{eqn1}). Using commercially-available finite element (COMSOL Multiphysics) and finite difference method (FIMMWAVE) software packages, the spatial variation of the electric field along a waveguide of any geometry can be calculated. For a waveguide as shown in \Fref{Fig1}, there is symmetry along the \textit{z}-direction, hence the electric field is translationally invariant in this direction resulting in the following form of the wave equation:

\begin{equation}
\label{eqn1}
[\nabla^{2}+n^{2}k^{2}]\vec{E}_t(x,y)= \beta^{2}\vec{E}_t(x,y).
\end{equation}

Here, $\vec{E}_t(x,y)$ is the transverse component of the electric field, \textit{n} is the refractive index, \textit{k} is the wavenumber and $\mathnormal\beta$ is the propagation constant. In the slot region, light propagates as two separate modes, one travelling in the fibre section at the top of the slot region and the other in the lower portion of fibre. The total field is written as:

\begin{eqnarray}
\label{eq2}
\fl\vec{E}(x,y)=\vec{E}_{1t}(x,y)\exp^{-\iota\beta_1z}+\vec{E}_{2t}(x,y)\exp^{-\iota\beta_2z}\nonumber\\=[\vec{E}_{1t}(x,y)+\vec{E}_{2t}(x,y)\exp^{-\iota(\beta_2-\beta_1)z}]\exp^{-\iota\beta_1z}
\end{eqnarray}

For cases where the two propagating modes have the same, or almost the same, effective index equation(\ref{eq2}) can be greatly simplified and this also allows us to neglect any mode beating effects since the mode beat length, $ L_B=2\pi / (\beta_2-\beta_1$), becomes much longer than the length of the slot cavity, \textit{L}$_s$. For a slot waveguide, the value of \textit{L}$_s$ will always be much lower than the beat length. In the following, we shall treat the solutions as super-modes, travelling with a single propagation constant, $\mathnormal\beta$, which has a value higher than either of the initial constants, $\mathnormal\beta_1$ and $\mathnormal\beta_2$, for symmetric modes. Moreover, we assume the two modes to be degenerate, which is true for a symmetric structure of this type. We can also consider the two modes given in Figures 2 (a) and (b) as being analogous to the two polarisation states of the $HE_{11}$ modes in a standard tapered optical fibre albeit with larger differences between their propagation constant.

\begin{figure}[htb!]
 \centering
 \includegraphics[scale=.4]{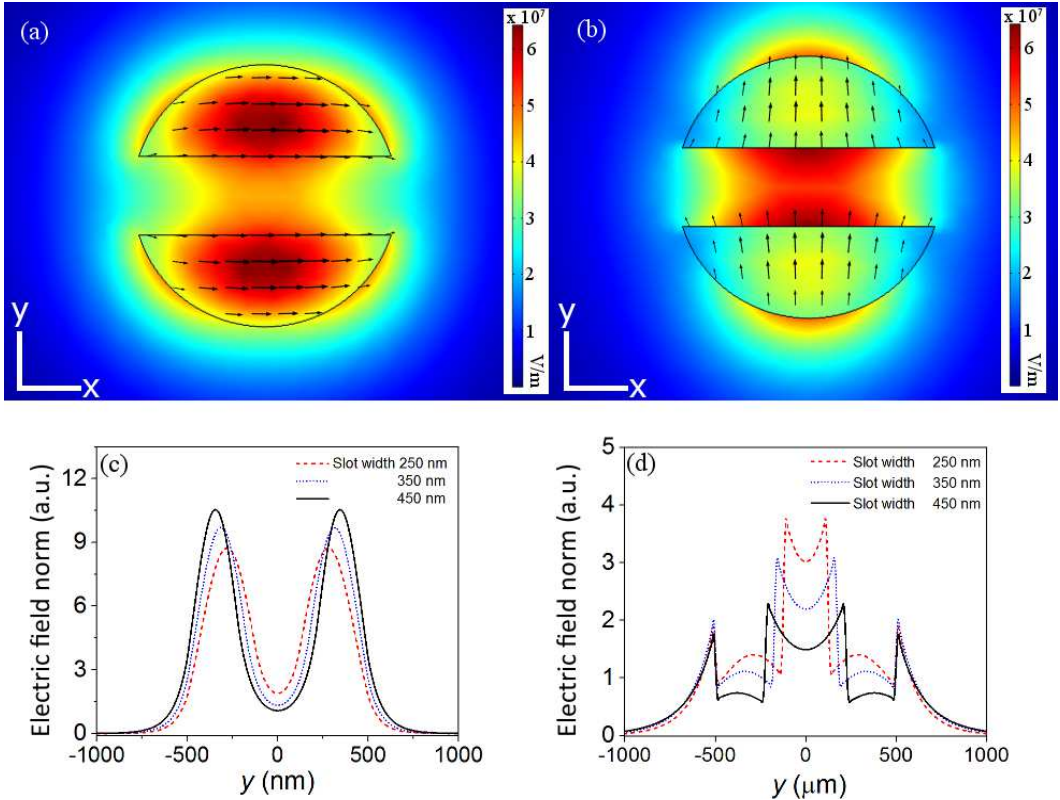}
 \caption{\label{Fig2}Electric field normal of modes generated in a 1 $\mu$m diameter fibre using (a) 720 nm light with parallel polarised light and (b) 1064 nm light with perpendicularly polarised light. (c) and (d) are the potentials along the \emph{y}-direction for (a) and (b), respectively. }
\end{figure}

\begin{figure}[htb!]
 \centering
 \includegraphics[scale=1]{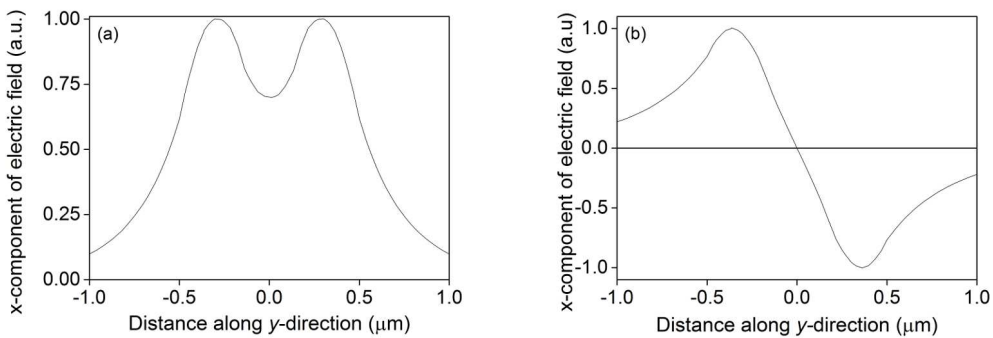}
 \caption{\label{Fig3}Symmetric and anti-symmetric modes for a single polarisation.}
\end{figure}

When the waveguide dimensions are comparable to the wavelength of light, the polarisations of the modes have more influence over the intensity distributions. This is most noticeable within the slotted region. Figures \ref{Fig2}(a) and (b) show two extreme cases where the electric field is polarised parallel and perpendicular to the slot walls. Figures \ref{Fig2}(c) and (d) give the electric field norms for parallel and perpendicular polarisations, respectively, and for slot widths, \textit{W}$_s$, varying from 250-450 nm. In the case of perpendicular polarisation, high intensities can be realized in the vacuum region between the slot walls since the component of the electric field normal to the boundary has no continuity requirement. Figure \ref{Fig3} shows the transverse field distributions corresponding to two modes for one polarisation at 1064 nm in a 1 $\mu$m diameter fibre with a 350 nm wide slot. \Fref{Fig3}(a) shows a symmetric mode, while \Fref{Fig3}(b) shows an anti-symmetric mode. All calculations throughout this paper are performed using the total electric field.
\subsection{Mode definition}
\begin{figure}[htb!]
 \centering
 \includegraphics[scale=1]{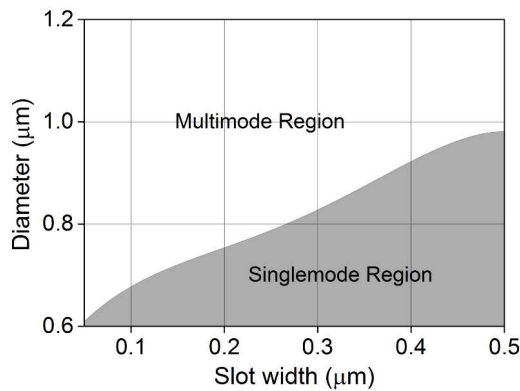}
 \caption{\label{Fig4}Graph of the regions of single-mode and multimode operation for 720 nm wavelength.}
\end{figure}

It has been shown that, in waveguides with a circular sector cross-section (i.e. a section of a circle which is enclosed by two radii and an arc) the mode numbers are non-integer, i.e.  $m=(p\pi)/\phi_0$ , where \textit{p} is an integer and $\phi_0$ is the sectoral angle \cite{ref31}. This non-integer mode number serves merely to indicate that full circular symmetry has been lost.  In this paper we are dealing with circular segments, as opposed to sectors. Hence, it is clear that the modes for this system will differ slightly, but one can assume that, in general, they will have similar behaviour. It is important to distinguish between what can be considered as \emph{single-mode} or \emph{multi-mode} in this system. In the absence of full analytical solutions, we define the single-mode regime as being the region where the only modes propagating are those which have intensity profiles with a single intensity maximum in each fibre segment as given in \Fref{Fig2}(a).   \Fref{Fig4} indicates the defined single and multimode regions for different fibre radii and slot widths as obtained from numerical models.

Higher-order modes can give rise to interesting intensity distributions within the slot and these could also be used for atom trapping. However, in this initial work we focus on the fundamental mode. To this end, we only consider parameters that are in, or near, the single-mode region, where contributions from the higher modes are either non-existent or small enough to be neglected.  It has been shown by Jung \textit{et al.} that modes can be effectively filtered out via selective excitation of the fundamental mode using a tapered fibre \cite{ref32}. Only symmetric fibre modes are considered as they are excited with much greater efficiency by the \textit{HE}$_{11}$ mode. Anti-symmetric modes should not be excited by an approximately uniform phase front. It can be seen from \Fref{Fig2}(a) that the parallel-polarised modes rapidly decay exponentially away from the slot walls. Therefore, these modes are better suited for blue-detuning so as to attract the atom to the intensity minimum at the centre of the slot. In contrast, the orthogonally polarised modes have a higher intensity than that of the parallel-polarised modes in the centre of the slot  (\Fref{Fig2}(b)) which causes atoms to be attracted towards the walls near the centre of the trap. As we shall see later, it is possible to combine these two field distributions with a combination of red- and blue-detuned light to draw atoms towards the centre of the slotted nanofibre.
\section{Trap design}
\subsection{Surface interaction potential}
A common issue in atom trapping near dielectric surfaces arises from the contribution to the total potential from the atom-surface interactions. Atoms near dielectric surfaces, such as the inner walls of the slot waveguide, are strongly affected by the attractive van der Waals (vdW) potential \cite{ref33, ref34, ref35}. To quantify this effect the Lennard-Jones (L-J) potential \cite{ref36} is often used as an approximation, such that

\begin{equation}
V_{vdW}=-\frac{C_3}{d^3},
\end{equation}
   where $d$ is the distance from the fibre surface to the atom and $C_3$ has the value of 3.362$\times 10^{-23}$ mK m$^3$. Assumptions that the major contribution to the atomic polarisability come from the first six lowest $P_{1/2}$ and $P_{3/2}$ levels of rubidium, and that the wall is a perfect conductor, have been made in this calculation \cite{ref37}. It has been shown \cite{ref38}  that instead of a full QED calculation of this potential a simple interpolation formula can be used. This formula agrees with the QED calculation to within 0.6$\%$. In a similar process we compare the L-J potential to this interpolation formula in an attempt to justify it, as it is used extensively throughout the literature as an approximation. We find that it agrees to within 1.4$\%$ at distances of less than 3 $\mu$m from the fibre surface.  Therefore, we use the L-J approximation throughout this paper.
\begin{figure}[htb!]
 \centering
 \includegraphics[scale=1.1
]{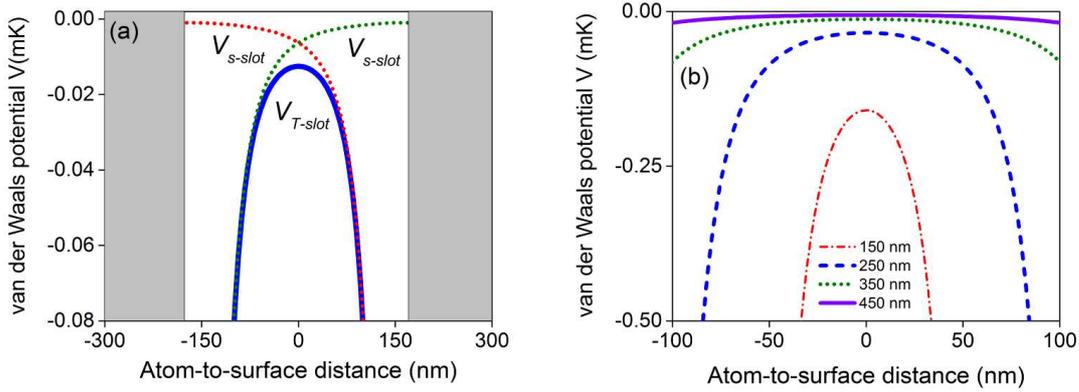}
 \caption{\label{Fig5}The van der Waals potential for an atom located in the fibre slot. (a) The total potential, \textit{V}$_{T-slot}$, is created by adding the contributions from each of the two walls, \textit{V}$_{s-slot}$ for a 1 $\mu$m fibre with a 350 nm slot width.  The fibre walls are indicated by the grey regions. (b) The total van der Waals potential as a function of atom-to-surface distance for $W_s$ values of 150 nm (dot-dash), 250 nm (dash), 350 nm (dot), and 450 nm (solid).}
\end{figure}
The effect of the van der Waals potential on neutral atoms located within the slot results in them being pulled towards the surface.  \Fref{Fig5}(a) indicates how the vdW potentials from either wall add to produce the total vdW potential seen by an atom in the slot. \Fref{Fig5}(b) explores how changing the slot width affects the total vdW potential. As atoms approach the slot walls, it is clear that the effect of the vdW potential becomes more prominent; also as $W_s$ becomes smaller atoms at the trap centre are affected more. The addition of a blue-detuned light field with respect to the atom transition frequencies alters this potential and creates a region in the centre with a stable equilibrium position.
\subsection{Optically produced potential and atom trapping}
A neutral atom interacting with an electric field $\vec{E}$ experiences a dipole potential given by
\begin{equation}
\label{eqn4}
U=-\frac{1}{4}\alpha(\omega)\vec{E}^*\vec{E},
\end{equation}

\begin{equation}
\label{eqn5}
\alpha(\omega)=\Sigma_nf_n[\frac{e^2/m}{\omega^2_n-\omega^2-\iota\omega\gamma_n}].
\end{equation}
where $\mathnormal\alpha$ is the atom polarisability as determined using Lorentz's model for a classical oscillator, \textit{e} is the electron charge, \textit{m} is the mass, $\mathnormal\omega_n$ is the natural frequency of the \emph{n}th oscillator, $\mathnormal\gamma_n$ is the damping coefficient of the \textit{n}th oscillator and \textit{f}$_n$ is the oscillator strength \cite{ref39}. It can be seen from equation (\ref{eqn5}) that the sign of the polarisability and, hence, the trapping potential is determined from the detuning of the laser fields involved. In the presence of an intensity gradient, the atom will experience a force along the gradient towards the intensity maximum in the case of red-detuned light, and towards the intensity minimum in the case of blue-detuned light. In order to trap atoms in the slot region we can choose symmetric modes that are polarised parallel or perpendicularly to the slot walls and detuned to the red or blue of the transition frequency.

Neutral atoms in the presence of a laser field of frequency $\mathnormal\omega$, which is close to an atomic resonance, experience a force which can be used to trap and even cool the atoms. As the laser frequency is detuned further and further from resonance, only the heating due to spontaneous scattering of this far-detuned laser field need be considered as the dominant mechanism for atom loss from the atom trap.  The atoms undergo a momentum recoil due to photon scattering. To quantify the usefulness of an atom trap the scattering rates should be determined and, hence, the trap lifetimes for atoms located near the centre of the trap. The scattering rate, $\mathnormal\Gamma_{sc}$ , for an atom in a dipole trap is given as \cite{ref37}:
\begin{equation}
\label{eqn6}
\Gamma_{sc}=\frac{3\pi c^2}{2\hbar \omega^3_0}\left(\frac{\Gamma}{\Delta}\right)^2I(r).
\end{equation}
Here, $c$ is the speed of light in vacuum, $\mathnormal\omega_0$ is the frequency at resonance, $\mathnormal\hbar$  is the reduced Planck's constant, $\Gamma$  is the dipole transition matrix element between the ground $|g\rangle$  and $|e\rangle$ excited states, and \textit{I(r)} is the intensity. Generally, one would have to solve for $\mathnormal{\Gamma}_{sc}$  by taking every atomic transition including the hyperfine structure into account. For the case of $^{87}$Rb in the \textit{D}$_{1/2}$ ground state, we can assume that the major contributions to this scattering rate are from the dipole transition rates from \textit{S}$_{1/2}$ to the \textit{P}$_{1/2}$ and \textit{P}$_{3/2}$ excited states. With these simplifications equation(\ref{eqn6}) becomes:
\begin{equation}
\label{eqn7}
\Gamma_{sc}=\frac{3\pi c^2}{2\hbar \omega^3_0}\left(\frac{\Gamma_{1/2}}{3\Delta_{1/2}}+\frac{2\Gamma_{3/2}}{3\Delta_{3/2}}\right)^2I(r),
\end{equation}
where $\mathnormal{\Gamma}_{1/2}$ and $\mathnormal{\Gamma}_{3/2}$ are the dipole transition matrix elements from the \textit{S}$_{1/2}$ to the \textit{P}$_{1/2}$ and \textit{P}$_{3/2}$ states, respectively.  The scattering rate will become more important later when we consider recoil heating losses due to atom scattering from the atom trap.
\subsection{Trapping potential}
While the proposed configuration has the capability of trapping rubidium, it should be clear that the methods laid out here, with the appropriate choice of dimensions and detunings, are transferable to the trapping of other neutral atom species.\Tref{tabone} lists the values chosen for the models used in this paper. Throughout the paper we refer to the blue-detuned power as \textit{P}$_b$ and the red-detuned power as \textit{P}$_r$.
\begin{table}[h]
\caption{\label{tabone}Parameters used in trapping potential models.}

\begin{indented}
\lineup
\item[]\begin{tabular}{@{}*{7}{l}}
\br
\ns
Parameter&Value\\
\mr
Blue-detuned wavelength $\lambda_b$&720 nm&\\
Red-detuned wavelength $\lambda_r$&1064 nm&\\
$C_3$ for $^{87}$Rb&3.362 J m$^3$\\
\br
\end{tabular}
\end{indented}
\end{table}

\subsubsection{Single colour slot trapping potential}

\begin{figure}[htb!]
 \centering
 \includegraphics[scale=.45]{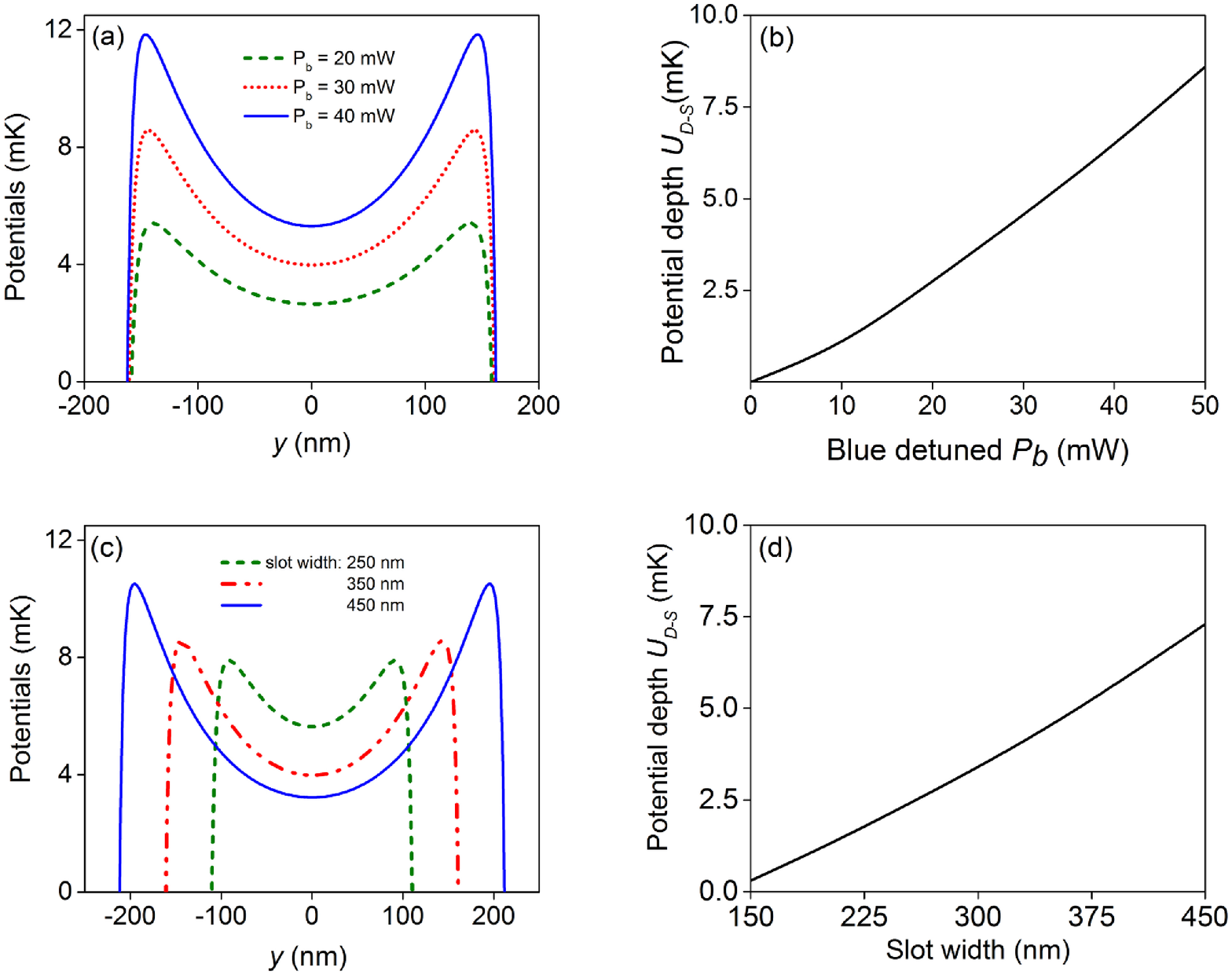}
 \caption{\label{Fig6}
(a) Combination of the blue-detuned potential and the van der Waals potential for a slot width of 350 nm. (b) Power \textit{P}$_b$ of the blue-detuned light vs. potential depth \textit{U}$_{total}$ for a slot width of 350 nm.  (c) Combination of the blue-detuned potential and the van der Waals potential for various slot widths and (d) dependence of trapping potentials on slot width. \textit{P}$_b$ is fixed at 30 mW.}
\end{figure}

Combining a blue-detuned light field with the vdW potentials arising from interactions of the atoms with both inner dielectric walls of the slotted fibre, a stable position in the \textit{y}-direction can be obtained. Unlike usual fibre-based atom traps which have potentials largely defined by the fibre radius, the proposed trap, in its most basic form, has two parameters, radius and slot width, both of which can be varied and are shown to have a large effect on the shape and efficiency of the atom potential, thereby giving more degrees of freedom in the design and increasing the flexibility of the system. \Fref{Fig6} illustrates how the potential can be altered by varying either of these two parameters. In the following sections the blue-detuned light is always chosen to be polarised parallel to the slot, i.e. along the \textit{x} direction.

Figures \ref{Fig6}(a) and (c) show the effect of changing blue-detuned powers and slot widths, respectively. By changing the power from 0-50 mW, or changing the slot width from 150-450 nm (\Fref{Fig6}(b) and (d)) trap depths up to 7.5 mK are obtainable. Unfortunately, this single colour trap provides no confinement along the \textit{x}-direction. To extend this idea into a stable trap in the \textit{xy}-plane, a red-detuned beam must be included, thereby creating a two-colour trapping scheme within the slot.

\subsubsection{Two-colour slot and fibre surface traps}

	When a second electric field, with sufficiently different frequency, is added, mode beating effects between the red- and blue-detuned modes can be neglected. We assume that the mode beating period between the two fields, \textit{E}$_r$ and \textit{E}$_b$, is much lower than the reaction time of atomic scale motion. Thus, we can assume that the two potentials add linearly such that the total potential is given by $U_T=U_b+U_r$. The addition of an attractive, red-detuned light field, polarised perpendicularly to the slot, creates a stable equilibrium position at the centre of the potential in both the \textit{x}- and \textit{y}-directions.
	
\begin{figure}[h!tb]
 \centering
 \includegraphics[scale=.21]{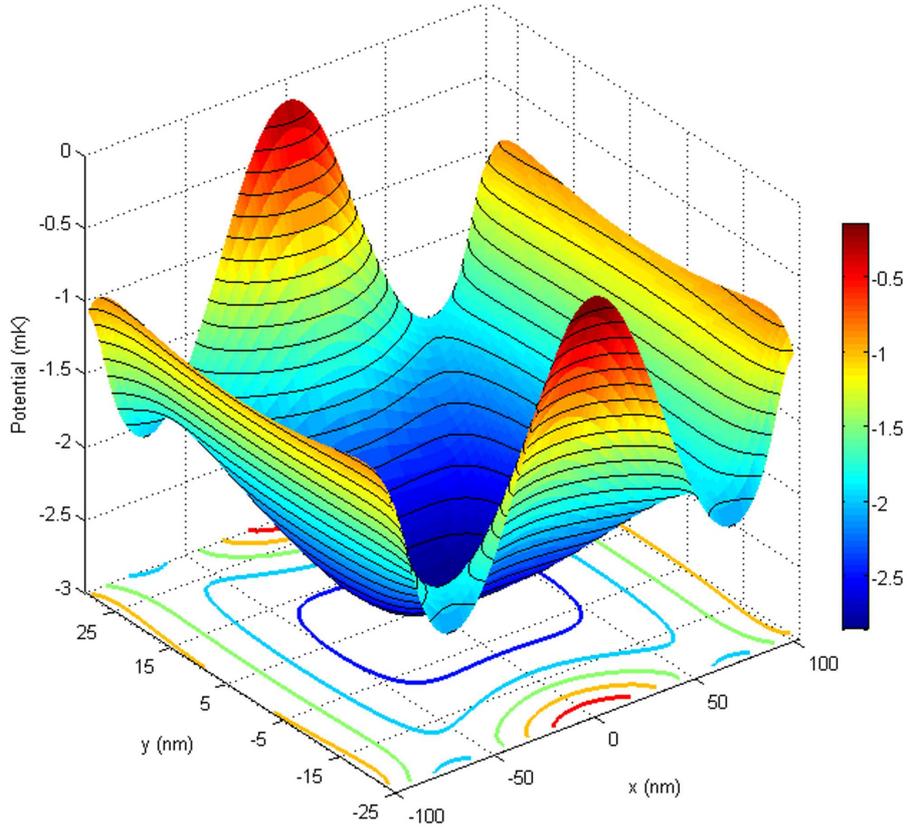}
 \caption{\label{Fig7}Combined surface and contour plot of a trapping potential in the \textit{xy}-plane for a 1 $\mu$m diameter fibre with a 350 nm slot width. \textit{P}$_r$=30 mW and \textit{P}$_b$=30 mW}.
 \end{figure}

 By adding the fields linearly, it can clearly be seen that a trapping well is formed in the \textit{xy}-plane (\Fref{Fig7}). The shape of this well varies dramatically in shape and depth with different choices of \textit{P}$_b$, \textit{P}$_r$, slot width, and fibre diameter.

 \begin{figure}[htb!]
 \centering
 \includegraphics[scale=1]{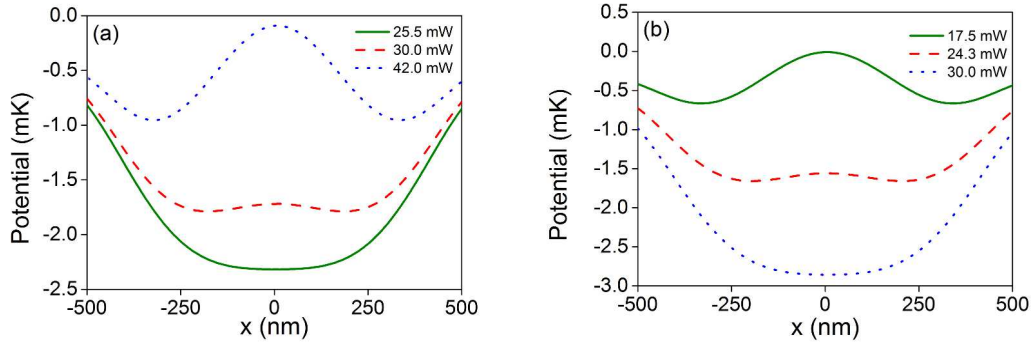}
 \caption{\label{Fig8}Two colour trapping potential for a 1 $\mu$m fibre with a 350 nm slot width in the \textit{x}-direction. (a) \textit{P}$_r$ is kept at 25 mW while \textit{P}$_b$ is varied. (b) \textit{P}$_b$ is fixed at 30 mW while \textit{P}$_r$ is varied.}
 \end{figure}

 The trap depth in the centre of the slot in the \textit{x}-direction is largely determined by the red-detuned power, \textit{P}$_r$. At the trap centre, the contribution from the vdW force is negligible and the contributions from the blue-detuned field are also low since the polarisation of the blue-detuned light is chosen to be parallel to the slot, thereby causing the power to decay away from the walls (see \Fref{Fig6}(a) and (c)). From Figure \ref{Fig8} we see that a change in the red-detuned power (a) causes the potential depth to change more rapidly than an equivalent change in power of the blue-detuned field (b). When the blue-detuned light field is sufficiently large compared to the red-detuned field the shape of the potential in the \textit{x}-direction veers away from being harmonic. To avoid complications, such a trap geometry is not considered when we determine trapping efficiencies.

 \begin{figure}[htb!]
 \centering
 \includegraphics[scale=1]{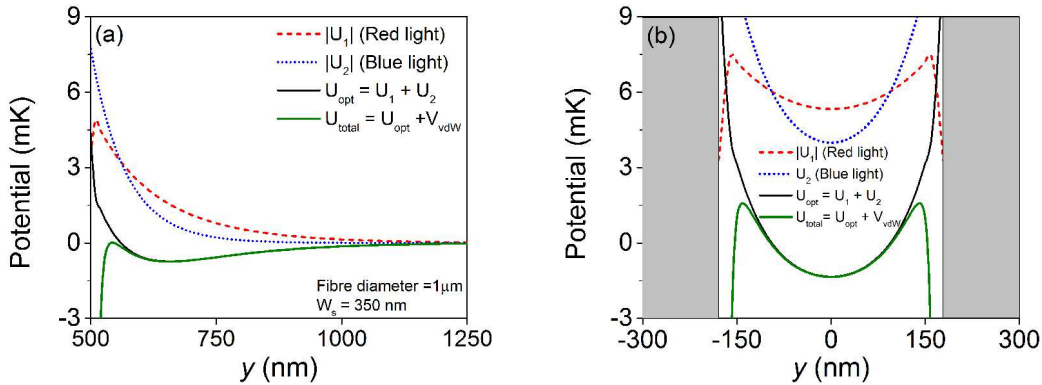}
 \caption{\label{Fig9}Contributions to the two-colour trap at (a) the outer fibre surface, and (b) the slot surfaces, to the total trapping potential \textit{U}$_{total}$. Both plots are taken in the \textit{y}-direction. The configurations are the same for (a) and (b), with \textit{P}$_b$= 25 mW and \textit{P}$_r$ = 30 mW. The grey areas in (b) represent the slot walls.}
 \end{figure}

 Varying the power ratio between the red and blue-detuned fields not only affects the form of the trapping potential inside the slot, \Fref{Fig9}(b), but also at the outer fibre surface, \Fref{Fig9}(a). The traps at the fibre surface resemble those proposed by Le Kien \textit{et al.} \cite{ref11}. An optimal trapping potential configuration for trapping atoms in the slot would maintain a deep trapping potential within the slot while keeping the trapping potential at the outer fibre surface at a local minimum, thereby preventing atoms from accumulating at the outer fibre surface due to the strongly attractive vdW potential.

\subsubsection{Trap optimisation}	

To optimise the trapping conditions we must consider four important parameters: slot width, fibre diameter, and the blue and red-detuned intensities resulting from the powers \textit{P}$_b$ and \textit{P}$_r$, respectively. Previously, we considered the confinement in the \textit{x}-and \textit{y}-direction. Here, only the field in the \textit{y}-direction is investigated for trap optimisation. This allows us to compare the fields both inside the slot region and at the outer fibre surface, whilst ensuring the vdW potential is correctly balanced within the slot region. It is known that the \textit{x}-direction potential forms a stable equilibrium point at the trap centre provided that one avoids the extreme situations such as those given in \Fref{Fig8}; therefore, only the \textit{y}-direction trapping potential need be used for determining an optimum configuration.

 \begin{figure}[htb!]
 \centering
 \includegraphics[scale=1]{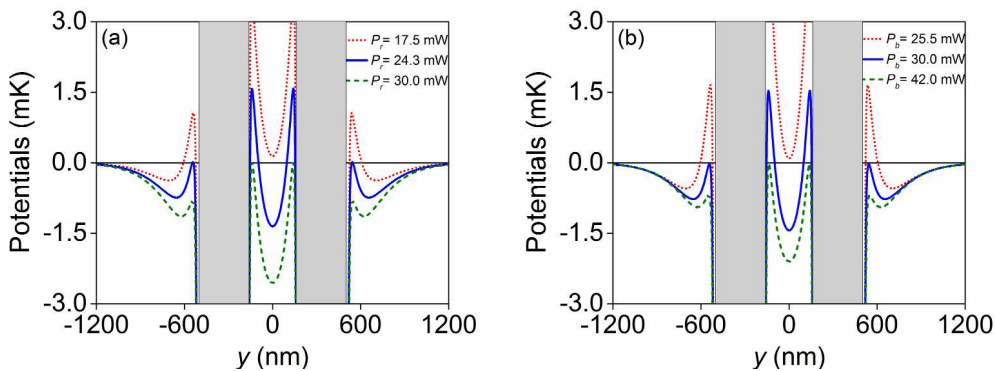}
 \caption{\label{Fig10}(a) Effect of the power \textit{P}$_r$ of the red-detuned light on total potential \textit{U}$_{total}$ while the power of blue-detuned light \textit{P}$_b$ is fixed at 30 mW; (b) Effect of the power \textit{P}$_b$ of blue-detuned light while \textit{P}$_r$ is fixed at 25 mW. All fibre parameters are the same as for \Fref{Fig9}.}
 \end{figure}

 By varying the input powers of the blue- and red-detuned beams, it is possible to choose an appropriate pair of values for use as default beam powers when considering the other parameters, such as slot width and fibre radius. The values are chosen in order to provide a deep potential inside the slot, with a minimum value below zero to facilitate the entry of rubidium atoms. For example, in Figures \ref{Fig10}(a) and (b), we can see that, within the slot region, the minimum value of the potential is only below zero for certain power configurations.

 \begin{figure}[htb!]
 \centering
 \includegraphics[scale=1.1]{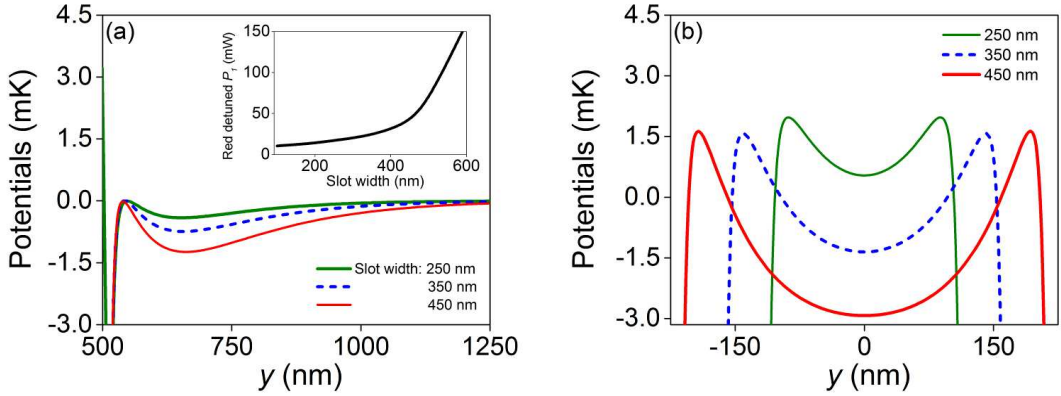}
 \caption{\label{Fig11} (a) Trap depth \textit{U}$_{D-F}$ at fibre surface and (b) trap depth \textit{U}$_{D-S}$ at slot fibre centre for different slot width values. The inset in (a) indicates the power \textit{P}$_r$ of red-detuned light required to keep the net optical potential depth \textit{U}$_{D-F}$ at a local minimum value for varying slot widths. The power of the blue-detuned light \textit{P}$_b$ is fixed at 30 mW and the fibre diameter is 1$\mu$m. }
 \end{figure}

 The power required to keep the outer surface trap depth at a local minimum as a function of slot width and $P_r$ is important in the choice of a suitable input beam power, leading to values in the 30 mW range. Motivation for this comes from two major facts: firstly, we can see from \Fref{Fig11} (a) that the power required to maintain the fibre surface trap at a minimum becomes unmanageable above a slot width of 400 nm and, secondly, we must constrain ourselves in some respect because there is no well-defined upper limit in a trap of this type. Higher powers lead to deeper traps, higher scattering rates and, eventually, fibre performance decay. A power of approximately 30 mW is  reasonable  for use with an optical nanofibre in ultrahigh vacuum for slot widths in the 250 nm to 450 nm range \cite{ref13,ref14}.

 \begin{figure}[htb!]
 \centering
 \includegraphics[scale=1]{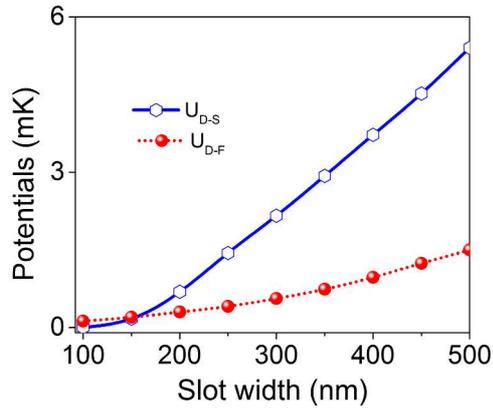}
 \caption{\label{Fig12} Trap depths \textit{U}$_{D-F}$ at fibre surface (dotted line) and \textit{U}$_{D-S}$ at slot fibre centre (solid line) versus slot width. The power of blue-detuned light \textit{P}$_b$ is fixed at 30 mW while the red-detuned light power \textit{P}$_r$ is varied to place the net optical potential depth \textit{P}$_{D-F}$ at a local minimum value (see the inset in \Fref{Fig11}(a)). }
 \end{figure}

 \Fref{Fig12} demonstrates how this choice of blue-detuned power can produce a trap with a large difference between the trap depths inside and outside the fibre. This also provides some motivation for the choice of a slot width of 350 nm. This value allows for a significant difference in the outer and inner potentials (\textit{U}$_{D-F}$ and \textit{U}$_{D-S}$, respectively) and this size is approximately one third of the diameter of the fibre, thereby alleviating most issues regarding the fibre's structural stability following etching.
  \begin{figure}[htb!]
 \centering
 \includegraphics[scale=1]{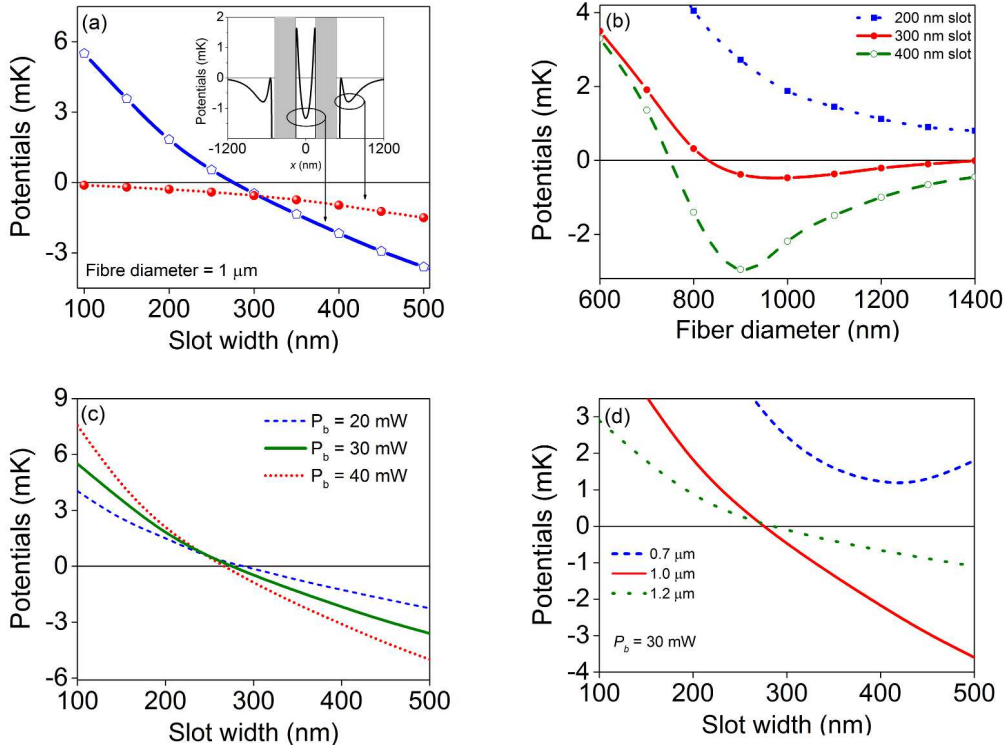}
 \caption{\label{Fig13}(a) Local minimum value of the trap depths at fibre surface \textit{U}$_{D-F}$ and at slot fibre center \textit{U}$_{D-S}$ versus slot width. For all Figures the powers of the red- and blue-detuned light, \textit{P}$_r$ and \textit{P}$_b$, are chosen to create an outer fibre surface trap depth at a local minimum value as in \Fref{Fig12}. (b) Minimum trap potential \textit{U}$_{D-S}$ vs fibre diameter for different slot widths, (c) minimum trap potential vs slot width for varying \textit{P}$_b$ powers, and (d) minimum trap potential vs slot width for various fibre diameters.}
 \end{figure}

 In an \emph{ad hoc} attempt to, at least locally, optimize the trapping conditions we begin by using a blue-detuned power, \textit{P}$_b$ = 30 mW, while varying the other parameters independently to see their effect on the potential. \Fref{Fig13}(b) gives a range of values where it is possible to obtain a suitable minimum potential value. Fibres of about 1 $\mu$m are located near the minimum of this curve for a 300 nm slot width. In \Fref{Fig13}(d) we show that the potential created using 30 mW of both red-detuned and blue-detuned power for both 1 $\mu$m and 1.2 $\mu$m fibres yield potentials that are below zero for slot widths greater than 275 nm. For a 0.7 $\mu$m fibre it is clear from the Figure that this condition is never met regardless of the chosen slot width. A 0.7 $\mu$m fibre still creates a strong trap with an adequate potential depth, but it lacks the capability of allowing atoms to easily enter the trap since atoms coming from infinity would need to overcome a net positive potential before entering the trap centre. As the fibre diameter is increased further, the mode will become well-confined to the larger segment sizes provided the slot widths are also in the same range as given in \Fref{Fig13}(d). This discussion does not preclude the use of smaller fibre diameters; it merely requires that the wavelengths of light chosen be tailored to the particular fibre size. A 0.7 $\mu$m diameter fibre, for instance, could be used provided a blue-detuned wavelength shorter than 720 nm is used.

Data in Figures \ref{Fig13}(b) and (d) indicate that higher slot widths produce better trap depths, but this comes at the cost of requiring higher red-detuned powers as is demonstrated by the inset of \Fref{Fig11}(a). Changing the blue detuned power has a relatively low effect on the potential as can be seen in \Fref{Fig13}(c). Fibre sizes much larger than 350 nm require very large red-detuned powers to compensate for the blue-detuned potential at the outer fibre surface. Our final choice of fibre is a 1 $\mu$m diameter fibre with a 350 nm slot width. The diameter and slot size were chosen by inspection of \Fref{Fig13}(b). Here, we  see that a low potential minimum occurs at this value. Larger slot sizes were investigated, but the power required to create a viable \textit{x}-direction trap began to exceed 40 mW once \textit{W}$_s>$ 350 nm slot widths were chosen.
\subsection{Atom trap viability}
In a two-colour trap the total scattering rate, $\mathnormal{\Gamma}_{total}$, is the sum of the scattering rates from the blue- and red-detuned fields, i.e. $\mathnormal{\Gamma}_{total}=\mathnormal{\Gamma}_{red}+\mathnormal{\Gamma}_{blue}$ . From this value, a characteristic coherence time $\mathnormal\tau_c=1/\mathnormal{\Gamma}_{total}$   can be determined \cite{ref39}. Each scattered photon from either field contributes some recoil energy to the atom. This will lead to a loss of atoms from the dipole trap. For a trap of depth $U_D$ a trap lifetime which only takes recoil heating into account for the chosen trap configuration is given as:
\begin{equation}
\tau_c=\frac{U_D}{2(E^r_{blue}\Gamma_{blue}+E^r_{red}\Gamma_{red})},
\end{equation}
where $E^r_{blue}$ and $E^r_{red}$ are the recoil energies associated with a blue and red photon, respectively. Along the \emph{x}-axis the contribution from the blue-detuned field is at its minimum value, hence the potential here reaches a maximum near or below zero. In contrast, the $y$-axis potential will have the same minimum value at the centre but, due to increasing blue-detuned intensity, as one approaches the walls the maximum value will be above zero due to the field's repulsive property. When determining the trap lifetimes we only consider the 1-D trap along the $x$-direction as it has a lower maximum depth than the $y$-direction.
\begin{table}[h]
\caption{\label{tabtwo}Trap parameters found by varying \textit{P}$_r$ with respect to \textit{P}$_b$, which was fixed at 30 mW. Values were obtained for a fibre width of 1 $\mu$m and a slot width of 350 nm. }

\begin{indented}
\lineup
\item[]\begin{tabular}{@{}*{7}{l}}
\br
\ns
Red-detuned power&Trap Depth \textit{U}$_D$&Scattering rate&Coherence time&Trap lifetime\\
(mW)&(mK)&(s$^{-1}$)&(s)&(s)\\
\mr
24.3&1.5&98.78&1.01E$^{-2}$&42\\
30&2.8&105.72&9.46E$^{-3}$&75\\
35&4&111.81&8.94E$^{-3}$&104\\
\br
\end{tabular}
\end{indented}
\end{table}

\section{Conclusions}
Optical micro and nanofibres have already been shown to be an invaluable tool in  trapping and probing  atomic systems. In this paper we propose that a nanostructured optical nanofibre is capable of trapping atoms inside the slot region, rather than at the surface, via the use of blue- and red-detuned beams. We investigated the use of the slotted waveguide structure over a broad range of configurations. By keeping the fibre diameters and slot widths near the curve given in \Fref{Fig4} and by using exponentially tapered fibres we can ensure that any contributions from the higher order modes is negligible \cite{ref40}. We have also shown that the polarisation of the light plays a crucial role in the realisation of such a trap. With the current electric field configuration, atoms will be trapped in line along the longitudinal axis of the structured fibre with potential minima located at distances of 140 - 200 nm from the slot walls. Using fundamental modes in the slotted region we have shown that a two-colour scheme is a viable method of producing adequate atom traps. For rubidium, trap depths of 4 mK, coherence times of the order of 10 ms and trap lifetimes of approximately 100 s should be accessible using modest input powers. Using an estimate for the optical depth per atom, namely the ratio of the on resonance scattering cross-section to the effective mode area $OD=\sigma_0/A_{eff}$, we obtain a value of 0.33 for the final configuration given in Table 2. Experimentally values of 0.08 have been measured \cite{ref13}. We deal exclusively with a slot which is assumed to have an infinite length, but in practice it would have a finite length and would be connected to an optical fibre pigtail at each end. Preliminary modelling work shows that $>70\%$ of the input mode can be transmitted beyond the end face of the slotted region. Flat end surfaces are the simplest case, but by tapering the slot ends we can significantly increase transmission into and out of the slot if required. Further analysis of this would require an in depth study of how the modes couple from the slot region back into the fibre modes and is beyond the scope of this paper.
		
This preliminary work opens up many other avenues for atom trapping, such as standing waves, two-colour traps, single-colour traps using higher order modes, magic wavelength traps, and the introduction of additional slots. We believe that, with this geometry, a stronger coupling of atom emissions into the guided modes of the waveguide should be achievable, as should more efficient interactions between the atoms and any other light fields present in the nanofibre due to the atom's close proximity to the slot walls. The scalability of this design, along with its long coherence times, high optical depth, and direct integration to optical systems make it an excellent candidate for quantum communication schemes. Experimentally, creating a slotted nanofibre is possible using a combination of existing techniques. Nanostructuring of an optical nanofiber using focused ion beam (FIB) techniques has recently been reported \cite{ref41}. Misalignment of the slot position along the fibre axis would lead to asymmetric potentials. A cursory investigation indicates that it would be possible to create a trapping potential for misalignments of the slot position from the fibre center of the order of 10 nm, which coincides with the accuracy of a FEI Helios Nanolab 650 Dualbeam FIB/SEM. Alternatively, processing of the optical fibre using tightly focussed \emph{fs} laser pulses could be considered \cite{ref42}.
\section{Acknowledgements}
This work is supported by OIST Graduate University.


\clearpage


\section*{References}
\begin{thebibliography}{10}
\bibitem{ref1}S F\'eron, J Reinhardt, S Le Boiteux, O Gorceix, J Baudon, M Ducloy, J Robert, Ch Miniatura, S Nic Chormaic, H Haberland and V Lorent 1993 Reflection of metastable neon atoms by a surface plasmon wave \textit{Optics communications} \textbf{102} 1
\bibitem{ref2}J P Dowling and J Gea-Banacloche 1996 Evanescent Light-Wave Atom Mirrors, Resonators, Waveguides, and Traps \textit{Advances In Atomic, Molecular, and Optical Physics} \textbf{36} 1
\bibitem{ref3}L Tong, R R Gattass, J B Ashcom, S He, J Lou, M Shen, I Maxwell and E Mazur 2003 Subwavelength-diameter silica wires for low-loss optical wave guiding \textit{Nature} \textbf{426} 816
\bibitem{ref4}A Yariv, 1985 Optical electronics, 3rd edition, Chapter 3, CBS College, New York.
\bibitem{ref5}S Ravets, J E Hoffman,L A Orozco, S L Rolston, G Beadie and F K Fatemi 2013 A low-loss photonic silica nanofiber for higher-order modes \textit{Optics Express} \textbf{21} 15 18325
\bibitem{ref6} L Russell, R Kumar, V B Tiwari and S Nic Chormaic 2013 Measurements on release-recapture of cold Rb-85 atoms using an optical nanofibre \textit{Opt. Commun.} \textbf{309} 313
\bibitem{ref7} L Russell, K Deasy, M Daly, M Morrissey and S Nic Chormaic 2012 Sub-Doppler temperature measurements of laser-cooled atoms using optical nanofibres \textit{Meas. Sci. Technol.} \textbf{23} 015201
\bibitem{ref8}K P Nayak, F Le Kien, M Morinaga, and K Hakuta Antibunching and bunching of photons in resonance fluorescence from a few atoms
into guided modes of an optical nanofiber 2009 \textit{Phys. Rev. A} 79, 021801
\bibitem{ref9} M J Morrissey, K Deasy, M Frawley, R Kumar, E Prel, L Russell, V G Truong, and S Nic Chormaic 2013 Spectroscopy, Manipulation and Trapping of Neutral Atoms, Molecules, and Other Particles Using Optical Nanofibers: A Review \textit{Sensors}  \textbf{13} 8 10449
\bibitem{ref10} B Ovchinnikov, S V Shul'ga and V I Balykin 1991 An atomic trap based on evanescent light waves \textit{J. Phys. B: At. Mol. Opt. Phys.} \textbf{24} 3173
\bibitem{ref11}F Le Kien, V I Balykin and K Hakuta 2004 Atom trap and waveguide using a two-color evanescent light field around a subwavelength-diameter optical fiber \textit{Phys. Rev. A} \textbf{70} 063403
\bibitem{ref12} F Le Kien, V I Balykin and K Hakuta 2005 State-Insensitive Trapping and Guiding of Cesium Atoms Using a Two-Color Evanescent Field around a Subwavelength-Diameter Fiber \textit{J. Phys. Soc. Japan} \textbf{74} 910
\bibitem{ref13}C Lacro\^{u}te, K S Choi, A Goban, D J Alton, D Ding, N P Stern and H J Kimble 2012 A state-insensitive, compensated nanofiber trap \textit{New J. Phys.} \textbf{14} 023056
\bibitem{ref14}J Lee, D H Park, S Mittal, M Dagenais and S L Rolston 2013 Integrated optical dipole trap for cold neutral atoms with an optical waveguide coupler \textit{New Journal of Physics} \textbf{15} 043010
\bibitem{ref15} E Vetsch, D Reitz, G Sage\'{e}, R Schmidt, S T Dawkins and A Rauschenbeutel 2010 Optical Interface Created by Laser-Cooled Atoms Trapped in the Evanescent Field Surrounding an Optical Nanofiber \textit{Phys. Rev. Lett.} \textbf{104} 203603
\bibitem{ref16} A Goban, K S Choi1, D J Alton, D Ding, C Lacro\^{u}te, M Pototschnig, T Thiele, N P Stern, and H J Kimble 2013 Demonstration of a State-Insensitive, Compensated Nanofiber Trap \textit{Phys. Rev. Lett.} \textbf{109} 033603
\bibitem{ref17}C L Hung, S M Meenehan, D E Chang, O Painter and H J Kimble 2013 Trapped atoms in one-dimensional photonic crystals \textit{New J. Phys.} \textbf{15} 083026
\bibitem{ref18}J D Thompson, T G Tiecke, N P de Leon, J Feist, A V Akimov, M Gullans, A S Zibrov, V Vuleti\'{c} and M D Lukin 2013 Coupling a Single Trapped Atom to a Nanoscale Optical Cavity \textit{Science} \textbf{340} 6137
\bibitem{ref19}M A Ol'Shanii, Yu B Ovchinnikov, V S Letokhov 1993 Laser guiding of atoms in a hollow optical fiber \textit{Opt. Commun.}  \textbf{98} 77
\bibitem{ref20}J P Burke, S Chu, G W Bryant, C J Williams, and P S Julienne 2002 Designing neutral-atom nanotraps with integrated optical waveguides \textit{Phys. Rev. A} \textbf{65} 043411
\bibitem{ref21}P Xu X He J Wang and M Zhan 2010 Trapping a single atom in a blue detuned optical bottle beam trap \textit{Opt. Lett.} \textbf{35} 2164
\bibitem{ref22} G Sagu\'{e} \textit{et al.} 2008 Blue-detuned evanescent field surface traps for neutral atoms based on mode interference in ultrathin optical fibres \textit{New J. Phys.} \textbf{10} 113008
\bibitem{ref23}C F Phelan, T Hennessy, and T Busch 2013 Shaping the evanescent field of optical nanofibers for cold atom trapping \textit{Opt. Express} \textbf{21} 27093
\bibitem{ref24}P Schneeweiss, F Le Kien, and A Rauschenbeutel 2014 Nanofiber-based atom trap created by combining fictitious and real magnetic fields \textit{New J. Phys.P} \textbf{16} 013014
\bibitem{ref25}F Le Kien and K Hakuta 2009 Microtraps for atoms outside a fiber illuminated perpendicular to its axis: Numerical results \textit{Phys. Rev. A} \textbf{80} 013415
\bibitem{ref26}M J Morrissey, K Deasy, Y Wu, S Chakrabarti, and S Nic Chormaic 2009 Tapered optical fibers as tools for probing magneto-optical trap characteristics \textit{Review of Scientific Instruments} \textbf{80} 053102
\bibitem{ref27}K P Nayak, F Le Kien, Y Kawai, K Hakuta, K Nakajima, H T Miyazaki, and Y Sugimoto 2011 Cavity formation on an optical nanofiber using focused ion beam milling technique \textit{Opt. Express} \textbf{19} 15 14040
\bibitem{ref28}L Zhang, J Lou, and L Tong 2011 Micro/nanofiber optical sensors \textit{Photonic Sensors} \textbf{1} 1 31
\bibitem{ref29} M C Frawley, A Petcu-Colan, V G Truong, S Nic Chormaic 2012 Higher order mode propagation in an optical nanofiber \textit{Optics Comm.} \textbf{285} 23 4648
\bibitem{ref30}P A Anderson, B S Schmidt and M Lipson 2006 High confinement in silicon slot waveguides with sharp bends \textit{Opt. Express} \textbf{14} 20
\bibitem{ref31}A Elsherbeni, D Kaifez, and S Zeng 1991 Circular Sectoral Waveguides \textit{IEEE Antennas and Propagation Magazine} \textbf{33} 6
\bibitem{ref32}Y Jung,Y Jeong, G Brambilla, and D J Richardson 2009 Adiabatically tapered splice for selective excitation of the fundamental mode in a multimode fiber \textit{ Opt. Lett.} \textbf{34} 15 2369
\bibitem{ref33}L Russell, D A Gleeson, V G Minogin, and S Nic Chormaic 2009 Spectral distribution of atomic fluorescence coupled into an optical nanofibre \textit{J. Phys. B: At. Mol. Opt. Phys.} \textbf{42} 185006
\bibitem{ref34} V G Minogin and S Nic Chormaic 2010 Manifestation of the van der Waals surface interaction in the spontaneous emission of atoms into an optical nanofiber \textit{Laser Phys.} \textbf{20} 2
\bibitem{ref35}M C Frawley, S Nic Chormaic, and V G Minogin 2012 The van der Waals interaction of an atom with the convex surface of a nanocylinder \textit{Phys. Scr.} \textbf{85} 058103
\bibitem{ref36}J E Lennard-Jones 1924 On the Determination of Molecular Fields \textit{Proc. R. Soc. Lond. A} \textbf{106} 738 463
\bibitem{ref37}A Derevianko, W R Johnson,  M S Safronova, and J F Babb 1999 High-Precision Calculations of Dispersion Coefficients, Static Dipole Polarizabilities, and Atom-Wall Interaction Constants for Alkali-Metal Atoms. \textit{Phys. Rev. Lett.} \textbf{82} 18 3859
\bibitem{ref38}R Marani, L Cognet, V Savalli, N Westbrook, C I Westbrook and A Aspect 2000 Using atomic interference to probe atom-surface interaction \textit{Phys. Rev. A} \textbf{61} 053402
\bibitem{ref39}R Grimm, M Weidem\"{u}ller and Yu B Ovchinnikov 2000 Optical dipole traps for neutral atoms. \textit{Ad. In At., Mole., and Opt. Phys.} \textbf{42} 95
\bibitem{ref40}A Petcu-Colan, M C Frawley and S Nic Chormaic 2011 Tapered few-mode fibers: mode evolution during fabrication and adiabaticity. \textit{JNOPM} \textbf{20} 293-307
\bibitem{ref41} M Daly, V G Truong, C Phelan and S Nic Chormaic 2013 Near-field trap for submicron particles and cold, neutral atoms using rectangular etched cavities in optical nanofibers. \textit{Frontiers in Optics 2013/Laser Science XXIX}
\bibitem{ref42} R Buividas, M Mikutis, G Gervinskas, D Day, G Slekys and S Juodkazis 2012 Femtosecond laser drilling of optical fibers for sensing in microfluidic applications  \textit{SPIE} \textbf{8463} Nanoengineering: Fabrication, Properties, Optics, and Devices IX 84630T







\end{thebibliography}
\end{document}